\documentclass[12pt]{article}

\usepackage{html,amsmath,amssymb,amsfonts,epsfig,cancel,cite,longtable,caption,siunitx,array,color,booktabs,setspace,arydshln}

\usepackage[T1]{fontenc}
\usepackage{dsfont}
\usepackage{mathrsfs}
\usepackage[mathscr]{euscript}
\usepackage{calligra}

\renewcommand\baselinestretch{1.2}
\numberwithin{equation}{section}

\newcommand{\be}{\begin{equation*}}
\newcommand{\ee}{\end{equation*}}
\newcommand{\beq}{\begin{equation}}
\newcommand{\eeq}{\end{equation}}
\newcommand{\ba}{\begin{array}}
\newcommand{\ea}{\end{array}}
\newcommand{\bea}{\begin{eqnarray}}
\newcommand{\eea}{\end{eqnarray}}
\newcommand{\bean}{\begin{eqnarray*}}
\newcommand{\eean}{\end{eqnarray*}}

\newcommand{\fref}[1]{Figure~\ref{#1}}

\newcommand{\capt}[3]{\parbox{#1}{\renewcommand{\baselinestretch}{1.2}
                                                           \caption{\label{#2}\small\it #3}}}

\newcommand{\IC}{\mathbb{C}}
\newcommand{\IP}{\mathbb{P}}

\newcommand{\IZ}{\mathbb{Z}}
\newcommand{\cO}{{\cal O}}
\newcommand{\cN}{{\cal N}}

\newcommand{\cA}{{\cal A}}
\newcommand{\cB}{{\cal B}}
\newcommand{\cC}{{\cal C}}
\newcommand{\cL}{{\cal L}}

\newcommand{\cV}{{\cal V}}

\def\cjn1{{\cA, \cC^*\otimes \wedge^j \cN^*}}
\def\bjn1{{\cA, \cB^*\otimes \wedge^j \cN^*}}
\def\vjn1{{\cA, \cV^*\otimes \wedge^j \cN^*}}
\def\cjn2{{\cA, \cC\otimes \wedge^j \cN^*}}
\def\bjn2{{\cA, \cB\otimes \wedge^j \cN^*}}
\def\vjn2{{\cA, \cV\otimes \wedge^j \cN^*}}

\newcommand{\cicy}[2]{\begin{matrix} #1\end{matrix}\!\left[\begin{matrix}#2 \end{matrix}\right]}

\begin{document}

\title{{\Large \bf$~$\\[-21pt]
Formulae for Line Bundle Cohomology \\ on Calabi-Yau Threefolds \\ 
}}

\vspace{3cm}

\author{
Andrei Constantin${}$ and Andre Lukas\\
}
\date{}
\maketitle
\thispagestyle{empty}
\begin{center} { 
	{\it 
       ${}^1$Department of Physics and Astronomy, Uppsala University, \\ 
       SE-751 20, Uppsala, Sweden\\[0.3cm]}
       
       {\it 
       ${}^1$Rudolf Peierls Centre for Theoretical Physics, Oxford University, \\ 
       Clarendon Laboratory, Parks Rd, Oxford OX1 3PU, United Kingdom\\[0.3cm]}
       }
\end{center}

\vspace{41pt}
\abstract
\noindent
We present closed form expressions for the ranks of all cohomology groups of holomorphic line bundles on several Calabi-Yau threefolds realised as complete intersections in products of projective spaces. The formulae have been obtained by systematising and extrapolating concrete calculations and they have been checked computationally. Although the intermediate calculations often involve laborious computations of ranks of Leray maps in the Koszul spectral sequence, the final results for cohomology follow a simple pattern. The space of line bundles can be divided into several different regions, and in each such region the ranks of all cohomology groups can be expressed as polynomials in the line bundle integers of degree at most three. The number of regions increases and case distinctions become more complicated for manifolds with a larger Picard number. We also find explicit cohomology formulae for several non-simply connected Calabi-Yau threefolds realised as quotients by freely acting discrete symmetries. More cases may be systematically handled by machine learning algorithms.

\vspace{40pt}
\noindent\rule{4cm}{0.4pt}

\noindent andrei.constantin@physics.uu.se, lukas@physics.ox.ac.uk

\newpage

\section{Introduction}
It is difficult to underestimate the importance of cohomology computations in mathematics and theoretical physics. Despite this, and except in simple cases, cohomology computations are hard to carry out explicitly. One situation where closed form expressions are known to exist is the case of line bundles on projective spaces. The result, known as Bott's formula, is strikingly simple: 
\begin{align*}
h^0(\IP^n, \cO_{\IP^n}(k)) &= \displaystyle{{k+n}\choose{n} } = \frac{1}{n!}\,(1+k)(2+k)\ldots(n+k)~, \text{ if }k\geq 0, \text{ and }0\text{ otherwise}.\\[4pt]
h^i(\IP^n, \cO_{\IP^m}(k)) &= 0~, \text{ if }0< i<n~.\\[8pt]
h^n(\IP^n, \cO_{\IP^n}(k)) &= \displaystyle{{-k-1}\choose{-n-k-1} } = \frac{1}{n!}\,(-n-k)\ldots(-1-k)~, \text{ if }k\leq -n-1, \text{ and }0\text{ otherwise}.
\end{align*}

An algorithm generalising Bott's formula to the case of toric line bundles has recently been proposed in Refs.~\cite{Blumenhagen:2010pv, Rahn:2010fm, Jow:2010}. This algorithm was developed in the context of string compactifications, where massless modes of the heterotic or type II string on compact Calabi-Yau manifolds are determined by vector bundle valued cohomology. However, passing from projective spaces or toric varieties to Calabi-Yau manifolds involves an additional layer of complication. Smooth Calabi-Yau manifolds can be realised as hypersurfaces or complete intersections in products of projective spaces or toric varieties. In the presence of such an embedding space,  knowledge about vector bundle cohomology on the embedding space can be transferred to the bundle restricted to the Calabi-Yau sub-manifold, using the Koszul complex and its associated spectral sequence. 

In general, cohomology computations with spectral sequences require explicit information about the ranks of the Leray maps. For complete intersection manifolds in products of projective spaces \cite{Candelas:1987kf, Candelas:1987du}, this information can be obtained using a computational algorithm that relies on the Bott-Borel-Weil theorem \cite{bestiary, Anderson:2008ex}. This algorithm has been implemented in Mathematica  \cite{cicypackage} and applied to various problems related to string compactifications \cite{Anderson:2007nc, Anderson:2008uw, Anderson:2009mh, Anderson:2011ns, Anderson:2012yf, Constantin:2013, Buchbinder:2013dna, Anderson:2013xka,  He:2013ofa, Buchbinder:2014qda, Buchbinder:2014sya, Buchbinder:2014qca, Anderson:2014hia, Constantin:2015bea, Buchbinder:2016jqr,  Braun:2017feb, Blesneag:2018ygh}. The experience gained by computing a large number of examples has led to the observation, made in Refs.~\cite{Constantin:2013,Buchbinder:2013dna}, that the ranks of cohomology groups of holomorphic line bundles on the tetra-quadric Calabi-Yau manifold follow a certain pattern that can be expressed by a concrete formula. 

The purpose of this note is to extend the above observation to several other complete intersection Calabi-Yau threefolds. We find formulae for line bundle cohomology for a number of other manifolds and this suggests that similar formulae may exist for large classes of manifolds. It is likely that they can be obtained systematically by making use of computer-aided learning techniques. 

There is at least one class of Calabi-Yau threefolds for which the appearance of closed form expressions for line bundle cohomology should not be surprising: smooth complete intersections in~$\IP^n$ with Picard number equal to one. Let $X\subset\mathbb{P}^n$ be such a manifold. All line bundles on $X$ can be obtained as restrictions of line bundles $\cL=\cO_{\IP^n}(k)$ on $\mathbb{P}^n$ and we denote these by $L = \cO_X(k)$. Their first Chern class can be written as $c_1(\cO_X(k))=kJ$, where $J$ is the restriction to $X$ of the K\"ahler form on $\mathbb{P}^n$. If $k>0$, Kodaira's vanishing theorem implies that $h^q(X,L)=0$, for all $q>0$. Hence $H^0(X,L)$ is the only non-trivial cohomology group and its rank equals the index of $L$. For negative line bundles the picture is reflected, due to Serre duality, $H^q(X,L)~=~H^{3-q}(X,L^\ast)$, which implies that $h^3(X,L)=-\text{ind}(L)$ and that all other cohomologies are trivial.
Using the Atiyah-Singer index theorem, the index of $L$ can be expressed as
\begin{equation*}
{\rm ind}(L) =\sum_{i=0}^3(-1)^i h^i(X,L)= \int_X \left( {\rm ch}_3 (L) + \frac{1}{12} c_2(TX) \wedge c_1(L)\right) = \frac{1}{6}d(X) k^3 + \frac{1}{12} d(X) \tilde c_2(TX) k~,
\end{equation*}
where $d(X)$ is the triple intersection number and $c_2(TX) = \tilde c_2(TX)\,J\wedge J$ is the second Chern class of $X$. Together with the information that $H^0(X,\cO_X) \simeq H^3(X,\cO_X)\simeq \IC$ and that $H^1(X,\cO_X)$ and $H^2(X,\cO_X)$ are trivial, this fixes the cohomology ranks for all holomorphic line bundles on complete intersection Calabi-Yau threefolds in $\IP^n$. There are five such manifolds, and the corresponding line bundle cohomology formulae were given in Ref.~\cite{Anderson:2008ex}:
\begin{equation*}
\begin{aligned}
h^0(\IP^4[\,5\,],\cO(k))&~=~{\rm Max}\left(\delta_{k,0}+  \frac{5}{6}\, k^3 + \frac{25}{6}\, k,~0\right)\\
h^0(\IP^5[\,3~3\,],\cO(k))&~=~{\rm Max}\left(\delta_{k,0}+ \frac{3}{2} k^3 + \frac{9}{2}\, k,~0\right)\\
h^0(\IP^5[\,4~2\,],\cO(k))&~=~{\rm Max}\left(\delta_{k,0}+   \frac{4}{3} k^3 + \frac{14}{3}\, k,~0\right)\\[4pt]
h^0(\IP^6[\,3~2~2\,],\cO(k))&~=~{\rm Max}\left(\delta_{k,0}+  2\, k^3 + 5\, k,~0\right)\\[4pt]
h^0(\IP^7[\,2~2~2~2\,],\cO(k))&~=~{\rm Max}\left(\delta_{k,0}+ \frac{8}{3}\, k^3 + \frac{16}{3}\, k,~0\right)~,\\
\end{aligned}
\end{equation*}
In addition, we have $h^1(X,{\cal O}(k))=h^2(X,{\cal O}(k))=0$ for all these manifolds and $h^3(X,{\cal O}(k))=h^0(X,{\cal }(-k))$ is obtained from the above results via Serre duality. We have used the notation commonly used in the physics literature by which, say,~$\IP^7[\,2~2~2~2\,]$ denotes a Calabi-Yau threefold embedded in $\IP^7$ and realised as the complete intersection of four hypersurfaces of degree~$2$.

We may encode the information contained in the above formulae in the following diagrams:
\begin{figure}[h]
\begin{center}
$h^0(X, \cO_X(k)):$~~~~~~ \raisebox{-0.14in}{\includegraphics[width=9cm]{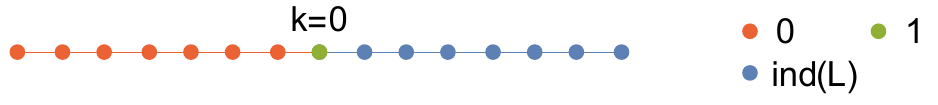}} 
$h^1(X, \cO_X(k)):$~~~~~~ \raisebox{-0.08in}{\includegraphics[width=9cm]{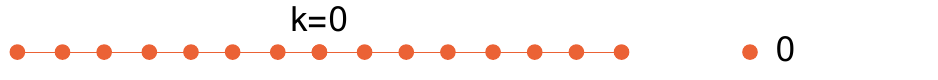}} \\[4pt]
$h^2(X, \cO_X(k)):$~~~~~~ \raisebox{-0.07in}{\includegraphics[width=9cm]{Pic1H1.pdf}} \\[8pt]
$h^3(X, \cO_X(k)):$~~~~~~ \raisebox{-0.16in}{\includegraphics[width=9cm]{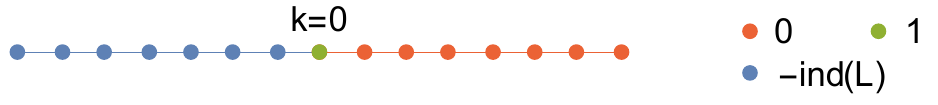}} \\[12pt]
\end{center}
\end{figure}

It is interesting to note that the same structure for the ranks of line bundle cohomology groups is present for threefolds with non-trivial canonical bundle, such as $\IP^3$. The difference comes from the fact that in this case Serre duality operates between cohomology groups $H^i(\IP^3, \cO_{\IP^3}(k))$ and $H^i(\IP^3, \cO_{\IP^3}(k-4))$.
\begin{figure}[h]
\begin{center}
\hspace{-19pt}$h^0(\IP^3, \cO_{\IP^3}(k)):$~~~~~~ \raisebox{-0.14in}{\includegraphics[width=8.3cm]{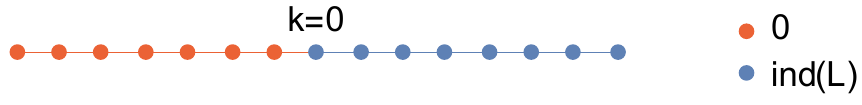}} 
$h^1(\IP^3, \cO_{\IP^3}(k)):$~~~~~~ \raisebox{-0.08in}{\includegraphics[width=9cm]{Pic1H1.pdf}} \\[4pt]
$h^2(\IP^3, \cO_{\IP^3}(k)):$~~~~~~ \raisebox{-0.07in}{\includegraphics[width=9cm]{Pic1H1.pdf}} \\[8pt]
\hspace{-12pt}$h^3(\IP^3, \cO_{\IP^3}(k)):$~~~~~~ \raisebox{-0.16in}{\includegraphics[width=8.6cm]{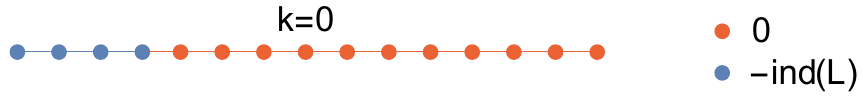}} \\[12pt]
\end{center}
\end{figure}

We will shortly turn to the case of manifolds with $h^{1,1}(X)>1$. Anticipating the results, we remark that the main features of the above formulae are retained in all the cases studied below. The ranks of all cohomology groups are given by simple expressions, although the intermediate calculations involving the Koszul resolution and the associated spectral sequence, kernels and co-kernels of Leray maps and so on are quite non-trivial. More concretely, we find that the ranks of all cohomology groups can be expressed as polynomials of degree at most three in the line bundle integers and the form of these polynomials changes in different regions of the $k$-space. 

\section{Manifolds with $h^{1,1}>1$}
Before discussing Calabi-Yau threefolds, it is interesting to take look at manifolds with Picard number greater than 1 for which line bundle cohomology formulae are known to exist. Products of projective spaces provide the simplest examples of such manifolds, and their line bundle valued cohomology can be obtained from Bott's formula combined with K\"unneth' formula
\begin{equation*}
H^i(\IP^{n_1} \times \IP^{n_2},\cO(k_1,k_2)) = \bigoplus_{i_1+i_2=i}H^{i_1}(\IP^{n_1},\cO(k_1))\otimes H^{i_2}(\IP^{n_2},\cO(k_2))~.
\end{equation*}

For concreteness, consider the line bundle $\cL = \cO_{\IP^1}(k_1)\otimes  \cO_{\IP^1}(k_2)$, with cohomology ranks given by the following formulae and illustrated in \fref{fig:P1P1}. Note that in this case there exist two lines, namely $k_1=-1$ and $k_2=-1$ along which all the cohomology ranks vanish\footnote{Such line bundles with entirely vanishing cohomology were recently studied for the case of toric varieties in \cite{Klaus:2018}.} and these lines separate the $k$-space into different regions in which the rank of one cohomology group does not~vanish.  
\begin{align*}	
h^0(\IP^1\times\IP^1, \cO(k_1,k_2))  &  = 
	\begin{cases}
	(1+k_1)(1+k_2)~, & k_1,k_2\geq0\\[12pt] 
	~0 & \text{otherwise}
	\end{cases}
\\[12pt]	
h^1(\IP^1\times\IP^1, \cO(k_1,k_2))  &  = 
	\begin{cases}
	(1+k_1)(-1-k_2)~, & k_1\geq 0,~k_2\leq -2\\[12pt] 
	(-1-k_1)(1+k_2)~, & k_1\leq-2,~k_2\geq 0\\[12pt]
	~0 & \text{otherwise}
	\end{cases}
\\[12pt]
h^2(\IP^1\times\IP^1, \cO(k_1,k_2))  &  = 
	\begin{cases}
	(-1-k_1)(-1-k_2)~, & k_1,k_2\leq-2\\[12pt] 
	~0 & \text{otherwise}
	\end{cases}\end{align*}

\begin{figure}[h]
\begin{center}
{\hskip-0pt
\begin{minipage}[t]{4.9in}
\vspace{-12pt}
{\includegraphics[width=6.7cm]{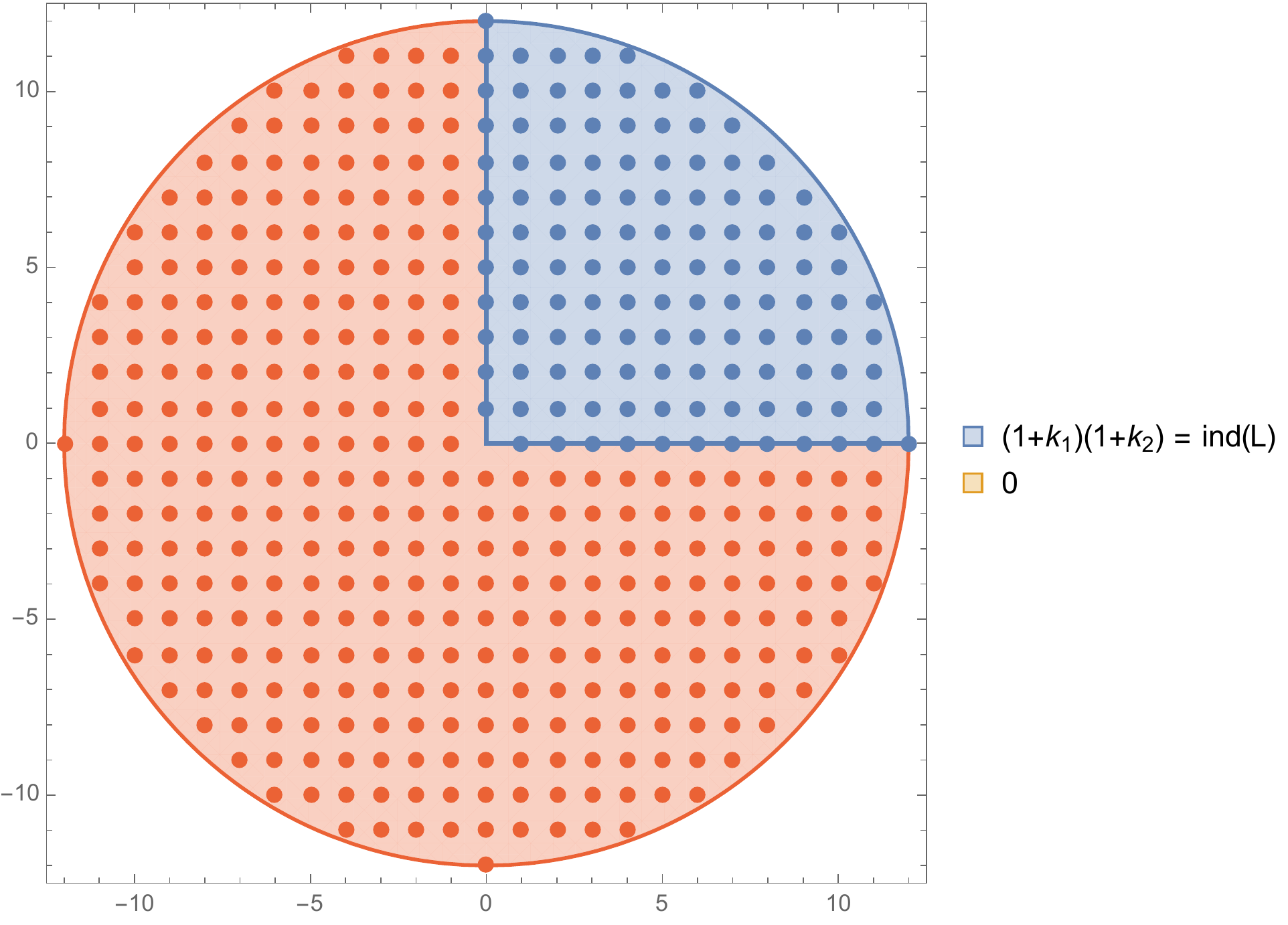}} 
\hfill \hspace*{14pt}
{\includegraphics[width=5.8cm]{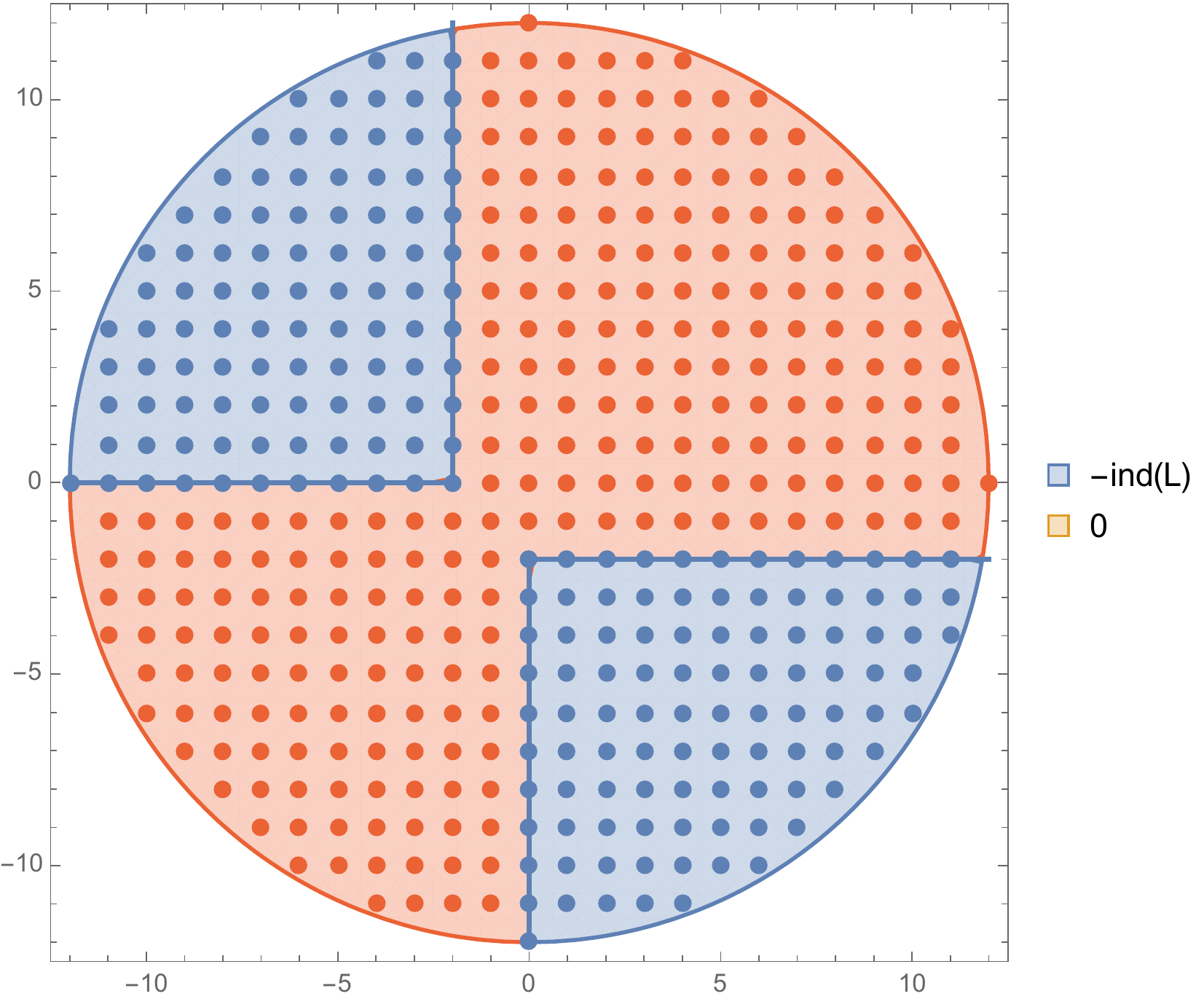}}
\hspace{0pt}
\vspace{10pt}
\end{minipage}}
\end{center}
\end{figure}
\vspace{-41pt}
\begin{figure}[h]
\begin{center}
{\hskip-0pt
\begin{minipage}[t]{4.9in}
\vspace{-22pt}
{\includegraphics[width=7.0cm]{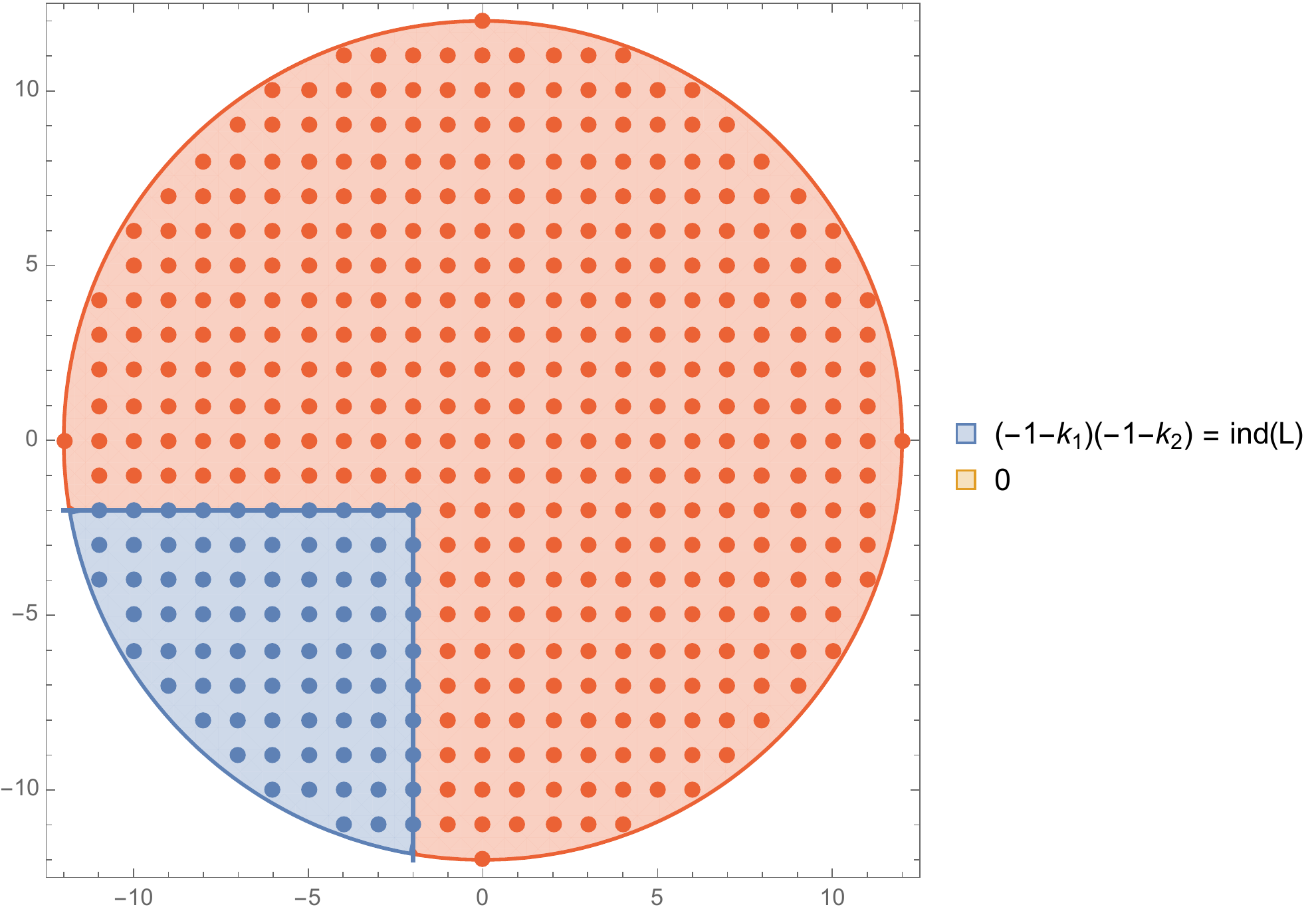}}
\hspace{0pt}
\vspace{10pt}
\end{minipage}}
\capt{6.4in}{fig:P1P1}{Regions in $k$-space where the cohomology ranks take different polynomial forms of degree at most $2$. In the blue regions the ranks are given by the index of the bundle, while in the red regions they vanish. Top left: $h^0(\IP^1\times\IP^1,\cL)$, top right: $h^1(\IP^1\times\IP^1,\cL)$, bottom: $h^2(\IP^1\times\IP^1,\cL)$.}
\end{center}
\end{figure}
\vspace{-21pt}

\subsection{Complete intersections in products of projective spaces}
Let $X\subset {\cal A}$ be a smooth complete intersection Calabi-Yau threefold in the ambient space  ${\cal A}:=\IP^{n_1}\times\dots\times\IP^{n_m}$, defined as the common zero locus of several multi-homogeneous polynomials. It is convenient to record the multi-degrees of the defining polynomials as a matrix, known as the configuration matrix, of the form
\begin{equation}
\label{conf}
\cicy{\IP^{n_1}\\[4pt] \vdots\\[4pt] \IP^{n_m}}{q^1_1&\cdots&q^1_R\\[4pt] \vdots&\hdots&\vdots\\[4pt]q^m_1&\hdots&q^m_R}^{h^{1,1}(X),~h^{2,1}(X)}
\end{equation}
There are $R$ defining polynomials, one for every column of the configuration matrix, and the integer vector ${\bf q}_a=(q_a^1,\ldots ,q_a^m)$, containing the entries of the $a^{\rm th}$ column of the above matrix, denotes the multi-degree of the $a^{\rm th}$ polynomial with respect to the homogeneous coordinates of the projective ambient space factors. The two non-trivial Hodge numbers $h^{1,1}(X)$ and $h^{2,1}(X)$ are attached as a superscript. Such a complete intersection $X$ has vanishing first Chern class if and only if the sum of the degrees in each row of the configuration matrix equals the dimension of the corresponding projective space plus one. 

For each projective factor we have an associated K\"ahler form and its restriction to $X$, which we denote by $J_r$, where $r=1,\ldots ,m$. The restrictions of the line bundles ${\cal L}={\cal O}_{\cal A}(k_1,\ldots ,k_m)=\cO_{\IP^{n_1}}(k_1)\otimes\ldots\otimes \cO_{\IP^{n_m}}(k_m)$ to $X$ are denoted by
$L = \cO_X(k_1,\ldots ,k_m)$, with first Chern classes $c_1( \cO_X(k_1,\ldots ,k_m))=\sum_r k^rJ_r$. In the cases discussed below, the second cohomology of $X$ is spanned by (the classes of) the forms $J_r$ so that all line bundles on $X$ are of the form  $\cO_X(k_1,\ldots ,k_m)$ and are classified by $m$-dimensional integer vectors $(k_1,\ldots ,k_m)$. We also refer to the integers $k_r$ as  ``line bundle integers''. With this notation the defining polynomials of $X$ are sections of the bundle ${\cal N}={\cal O}_{\cal A}({\bf q}_1)\oplus\cdots\oplus {\cal O}_{\cal A}({\bf q}_R)$ and $R = {\rm rank}(\cN)$.

The Atiyah-Singer index theorem applied to such line bundles now leads to
\begin{equation*}
{\rm ind}(L)=\sum_{i=0}^3(-1)^i h^i(X,L)=\int_X \left( {\rm ch}_3 (L) + \frac{1}{12} c_2(TX) \wedge c_1(L)\right) =
\frac{1}{6}d_{rst}k^rk^sk^t+\frac{1}{12}c_2^rk_r\; ,
\end{equation*}
where $d_{rst}=\int_XJ_r\wedge J_s\wedge J_t$ are the triple intersection numbers and $c_2^r=\int_X c_2(TX)\wedge J_r$ are the components of the second Chern class of $X$. Summation over repeated indices is understood. This still provides one easy-to-compute relation between the four cohomology ranks but, unlike in the Picard number one case, there are now line bundles $L$ (other than the trivial bundle) such that neither $L$ nor $L^*$ is ample. For these line bundles no additional information from Kodeira's vanishing theorem is available and detailed computations of the cohomology groups are required. Such computations are usually based on the Koszul sequence
\begin{equation}
 0~\longrightarrow~ \wedge^R{\cal N}^*\otimes{\cal L}~\longrightarrow~  \wedge^{R-1}{\cal N}^*\otimes{\cal L}~\longrightarrow\cdots\longrightarrow~ {\cal N}^*\otimes{\cal L}~\longrightarrow~{\cal L}~\longrightarrow~ L~\longrightarrow~ 0\; , \label{Koszul}
\end{equation} 
combined with spectral sequence methods, the Bott-Borel-Weil result for cohomologies on the ambient space ${\cal A}$ and, in many cases, knowledge of the maps involved. These methods, implemented in Ref.~\cite{cicypackage}, have been used for the specific calculations on which our results below are based. The intermediate steps in those calculations can be complicated, typically the more so the higher the co-dimension $R$. As such, there seems to be no a-priori reason for the ranks of line bundle cohomology groups to be simple when expressed in terms of the line bundle integers~$k_r$. However, our results below indicate that they are. Note, since $H^3(X,L)\cong H^0(X,L^*)$ and $H^2(X,L)\cong H^1(X,L^*)$ by Serre duality, it is sufficient to present the results for $h^0(X,L)$ and $h^1(X,L)$ for all line bundles $L$.

Our approach is empirical. For a given manifold $X$ we compute the cohomology for a large number of line bundles ${\cal O}_X(k_1,\ldots ,k_m)$ using the code in Ref.~\cite{cicypackage}. From these results we identify a number of regions in $k$-space and, for each such region, a cubic polynomial in the integers $k_r$ which describe the cohomology ranks in this region. The number of explicit cohomologies computed is significantly larger than the number of coefficients in the Ansatz, so there is strong evidence the formulae are correct. The formulae presented below have been checked for all line bundles with integers in the range $-10\leq k_r\leq 10$, and in same cases for many more. 

We should add a word of caution. A configuration matrix~\eqref{conf} really describes a family of manifolds, parametrised by the complex structure moduli which are encoded in the coefficients of the defining polynomials. Line bundle cohomology ranks have generic values in this moduli space but it is also known that they can jump at specific, non-generic loci, due to the complex structure dependence of the maps in the sequence~\eqref{Koszul}.  All the cohomology results presented in this note are valid for generic choices of the defining polynomials. Investigating the situation at jumping loci would be interesting but goes beyond our present scope.

\subsection{The bicubic manifold}

Let $X$ be a generic threefold in the ambient space ${\cal A}=\mathbb{P}^2\times\mathbb{P}^2$ defined by the configuration matrix
\beq
\cicy{\IP^2 \\ \IP^2}{\,3~ \\ \,3~}^{2,83}~
\eeq
and $L = \cO_X(k_1,k_2)$ a line bundle over $X$. Due to the symmetry of this configuration we have $h^q(\cO_X(k_1,k_2))=h^q(\cO_X(k_2,k_1))$, so without loss of generality we can assume that $k_1\leq k_2$. The corresponding cohomology formulae are given below, together with two plots in~\fref{fig:Bicubic} showing the regions where the expressions take different forms.

\begin{align}
\label{H0:bicubic}
h^0(X,L)  &  = 
	\begin{cases}
	\displaystyle \frac{1}{2}(1+k_2)(2+k_2)~, & k_1=0,~k_2\geq 0\\[8pt] 
	{\rm ind}(L)~, & k_1,k_2>0\\[8pt]
	0 & \text{otherwise}
	\end{cases}
\\[8pt]
\label{H1:bicubic}
h^1(X,L)  & = 
	\begin{cases}
	\displaystyle \frac{1}{2} (-1+ k_2) (-2 + k_2) ~, & k_1=0,~k_2>0 \\[8pt] 
	\displaystyle-{\rm ind}(L)~, & k_1<0,~k_2>-k_1\\[8pt]
	0 & \text{otherwise}~,
	\end{cases}
\end{align}
Here, the index is explicitly given by $\displaystyle {\rm ind}(L)=\frac{3}{2} (k_1 + k_2) (2 + k_1 k_2)$. 

\begin{figure}[h]
\begin{center}
{\hskip0pt
\begin{minipage}[t]{5.4in}
\raisebox{0in}{\includegraphics[width=7.2cm]{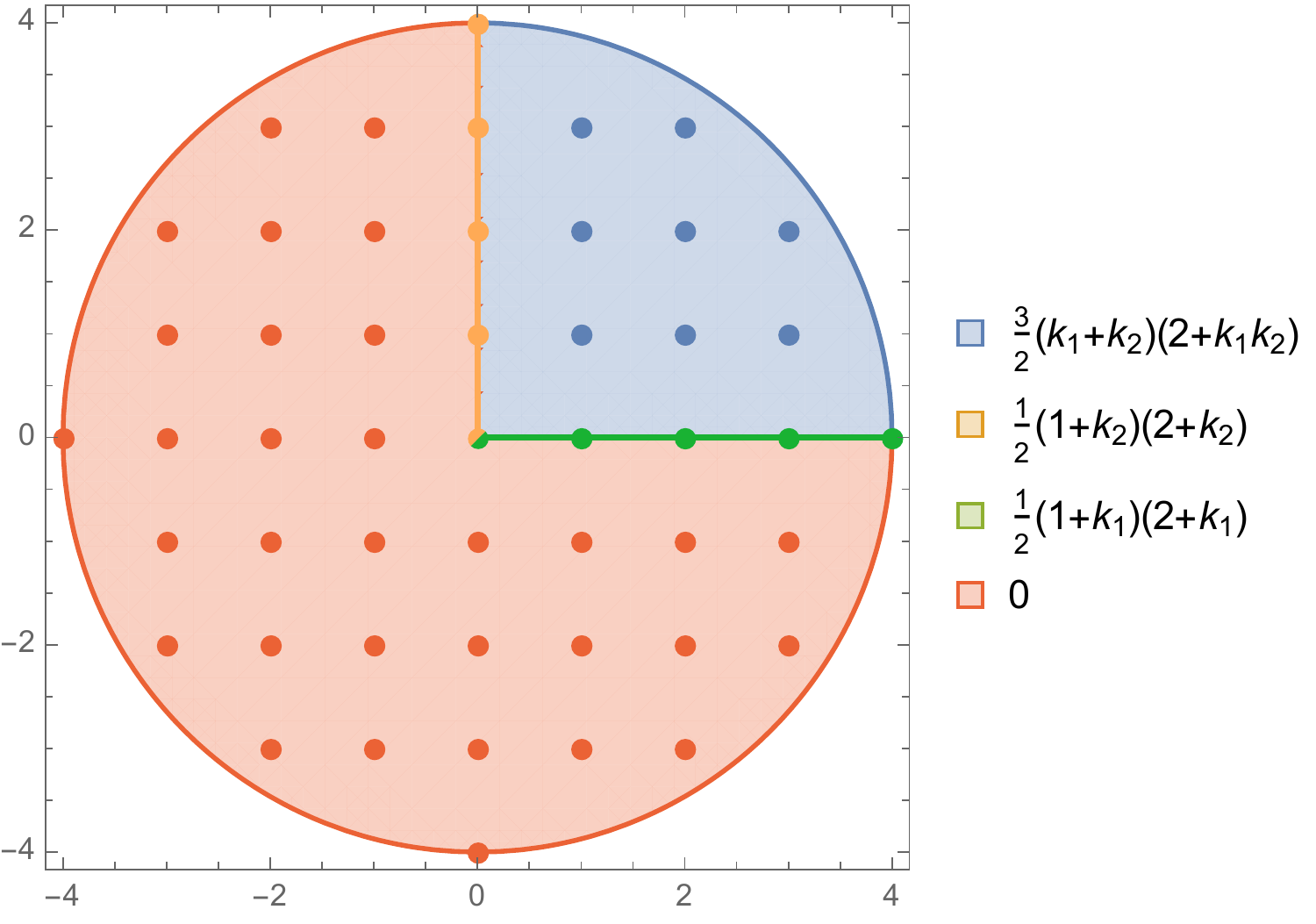}} 
\hfill \hspace*{14pt}
\raisebox{0in}{\includegraphics[width=7.2cm]{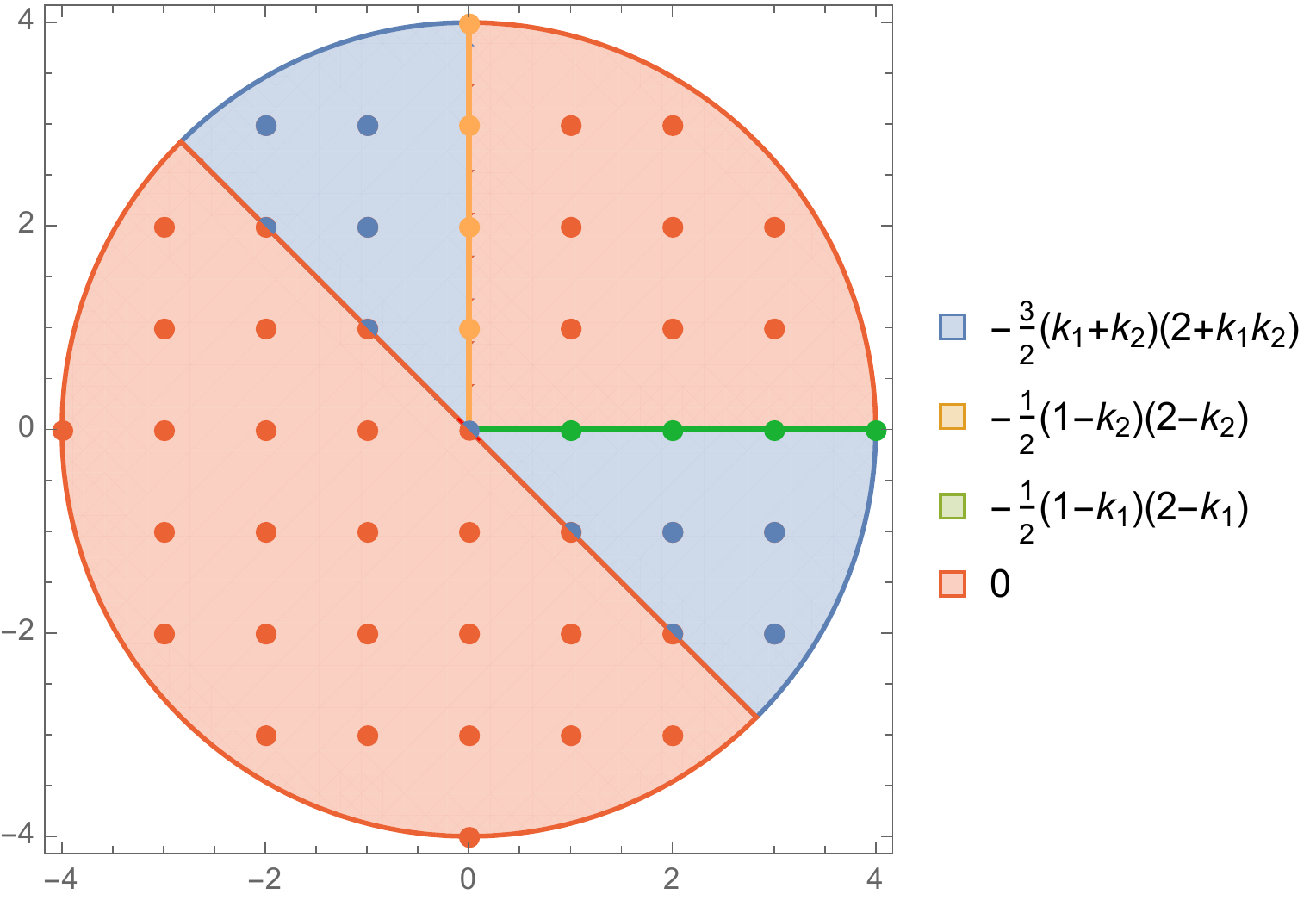}}
\hspace{10pt}
\end{minipage}}
\capt{6.4in}{fig:Bicubic}{Regions in $k$-space where $h^0(X,L)$ (left) and $h^1(X,L)$ (right) take different polynomial forms. In the blue regions $h^0(X,L) = {\rm ind}(L)$ and $h^1(X,L) = -{\rm ind}(L)$.  By Serre duality, the plots for $h^2(X,L)$ and $h^3(X,L)$ are obtained from the plots for $h^1(X,L)$ and, respectively, $h^0(X,L)$ by reflection about the origin.}
\end{center}
\end{figure}

The expressions given above for the semi-axis $k_1=0,k_2> 0$ and, implicitly by symmetry, for the semi-axis $k_1>0, k_2=0$ can be combined into the single formula
\begin{equation*}
\begin{aligned}
h^0(X,L) &~=~ \displaystyle \frac{1}{4}(1+k_1)(2+k_1)(1+k_2)(2+k_2)~,\\[4pt]
h^1(X,L) &~=~ \displaystyle \frac{1}{4}(-1+k_1)(-2+k_1)(-1+k_2)(-2+k_2)~,\\[4pt]
h^2(X,L) &~=~ h^3(X,L) ~=~ 0~.
\end{aligned}
\end{equation*}
Computing the characteristic, one recovers the formula for the index, as expected.

The above results have been inferred from the results of explicit cohomology calculations for many values of $k_1$, $k_2$, using the computer code~\cite{cicypackage}.  However, the bicubic manifold is sufficiently simple so that we can derive these formulae with relative ease by either combining vanishing theorems with the index or else from the sequence~\eqref{Koszul}. For $k_1>0$ and $k_2>0$, Kodaira's vanishing theorem ensures that $H^0(X,L)$ is the only non-trivial cohomology group, and its rank equals the index of $L$.
 Similarly, when $k_1<0$ and $k_2<0$, the only non-trivial cohomology is $H^3(X,L)$. Hence we only need to study the points lying in the second and fourth quadrant, as well as the points lying along the lines $k_1=0$ and $k_2=0$. For this, we can make use of the embedding of $X$ in $\IP^2\times \IP^2$. In fact, due to the symmetry of the problem, it suffices to study only the points lying in the second quadrant and on its boundary.
 
For the bicubic the Koszul sequence specialises to
\begin{equation*}
0 ~\longrightarrow~ \cL \otimes \cN^*  ~\stackrel{p}{\longrightarrow}~ \cL ~\longrightarrow~ \cL|_X ~\longrightarrow~ 0~,
\end{equation*}
where $\cN = \cO_{\cal A}(3,3)$ and $p$ is the defining polynomial. Passing to cohomology, we have the following long exact sequence: 
\begin{equation}
\label{cohsequence:bicubic}
\begin{aligned}
0 ~ &\longrightarrow~ H^0(\cA,\cL \otimes \cN^*)  ~\longrightarrow~ H^0(\cA, \cL) ~\longrightarrow~ H^0(X,\cL|_X) ~\longrightarrow~ \\[4pt]
&\longrightarrow~ H^1(\cA,\cL \otimes \cN^*)  ~\longrightarrow~ H^1(\cA, \cL) ~\longrightarrow~ H^1(X,\cL|_X) ~\longrightarrow~ \\[4pt]
&\longrightarrow~ H^2(\cA,\cL \otimes \cN^*)  ~\longrightarrow~ H^2(\cA, \cL) ~\longrightarrow~ H^2(X,\cL|_X) ~\longrightarrow~ \\[4pt]
&\longrightarrow~ H^3(\cA,\cL \otimes \cN^*)  ~\longrightarrow~ H^3(\cA, \cL) ~\longrightarrow~ H^3(X,\cL|_X) ~\longrightarrow~ 0
\end{aligned}
\end{equation}
hence $H^q(X,L)$ for $q=0,1,2$ is given by
\begin{equation*}
\begin{aligned}
H^q(X,L)~\simeq~{\rm Coker} & \left(H^q(\cA,\cL \otimes \cN^*)  ~\stackrel{p}{\longrightarrow}~ H^q(\cA, \cL) \right) ~\oplus\\[4pt]
{\rm Ker} & \left( H^{q+1}(\cA,\cL \otimes \cN^*)  ~\stackrel{p}{\longrightarrow}~ H^{q+1}(\cA, \cL) \right)\; .
\end{aligned}
\end{equation*}
Along the semi-line $k_1=0$, $k_2\geq 0$, the only non-trivial bundle-valued cohomology groups on ${\cal A}=\IP^2\times \IP^2$ that appear in the long exact sequence \eqref{cohsequence:bicubic} are $H^0(\cA, \cL)$ and $H^2(\cA,\cL\otimes \cN^*)$. Hence $H^0(X,L)\simeq H^0(\cA, \cL)$, $H^1(X,L)\simeq H^2(\cA, \cL\otimes \cN^*)$. The corresponding ranks can be obtained using Bott's formula, and this leads to the corresponding results given in \eqref{H0:bicubic} and \eqref{H1:bicubic}. Also, it follows that $H^2(X,L)$ and $H^3(X,L)$ are trivial. 
Due to the symmetry of the configuration, the semi-line $k_1\leq 0$, $k_2 = 0$ is Serre dual to the semi-line $k_1=0$, $k_2\geq 0$ and hence $H^0(X,L)$ and $H^1(X,L)$ are trivial in this case.

Finally, we have the situation $k_1<0$, $k_2>0$. In this case, the only non-trivial cohomologies on ${\cal A}=\IP^2\times \IP^2$ are  $H^2(\cA,\cL)$ and $H^2(\cA,\cL\otimes \cN^*)$. It follows that $H^0(X,L)$ and $H^3(X,L)$ are trivial and 
\begin{equation}
\label{H1ker:bicubic}
\begin{aligned}
H^1(X,L)&~\simeq~
{\rm Ker} \left( H^2(\cA,\cL \otimes \cN^*)  ~\stackrel{p}{\longrightarrow}~ H^2(\cA, \cL) \right)~\\[4pt]
H^2(X,L)&~\simeq~
{\rm Coker} \left( H^2(\cA,\cL \otimes \cN^*)  ~\stackrel{p}{\longrightarrow}~ H^2(\cA, \cL) \right)~.
\end{aligned}
\end{equation}
The rank of the map $p$ that appears in \eqref{H1ker:bicubic} turns out to be always maximal which can be shown by methods of commutative algebra. Hence, for the line bundles for which $h^2(\cA,\cL \otimes N^*)\leq h^2(\cA, \cL)$, we conclude that $H^1(X,L)$ is trivial, while $h^2(X,L)$ must equal ${\rm ind}(L)$. Similarly, when $h^2(\cA,\cL \otimes N^*)> h^2(\cA, \cL)$, $H^2(X,L)$ is trivial and $h^1(X,L)= -{\rm ind}(L)$. The boundary between these two phases is given by ${\rm ind}(L) = 0$, which corresponds to the line $k_2=-k_1$. 

\subsection{Another hypersurface with Picard number two}

Let $X$ be a generic member of the family of threefolds defined in the ambient space ${\cal A}=\mathbb{P}^1\times\mathbb{P}^3$ by the configuration matrix
\beq
\cicy{\IP^1 \\ \IP^3}{\,2~ \\ \,4~}^{2,86}~
\eeq
and $L = \cO_X(k_1,k_2)$ a line bundle over $X$.  From our explicit cohomology calculations for may values of $k_1$, $k_2$ we infer the following formulae:

\begin{align}
h^0(X,L)  & = \begin{cases}\displaystyle k_1+1~, & k_1\geq0,~k_2=0 \\[8pt] 
\displaystyle {\rm ind}(L), & k_1\geq0,~k_2>0\\[8pt]
-k_1+1,~ & k_1<0~,k_2=-4k_1\\[8pt]
\displaystyle\frac{32}{3} k_1(1 - k_1^2)  +  {\rm ind}(L) ,~ & k_1<0~,k_2>-4k_1\\[8pt]
0&\text{otherwise}\end{cases}\\[12pt]
\vspace{21pt}
h^1(X,L)  &= \begin{cases}-(k_1+1)~, & k_1<0,~k_2=0\\[8pt]
\displaystyle- {\rm ind}(L) , & k_1<-1,~-4k1>k_2>0  \\[8pt] 
\displaystyle -k_1+1 - {\rm ind}(L)~,& k_1 \leq -1,~ k_2 = -4 k_1\\[8pt] 
\displaystyle \frac{32}{3}   k_1 (1 - k_1^2)~,& k_1 \leq -1,~ k_2 > -4 k_1\\[8pt] 
0 & \text{otherwise}\end{cases}
\end{align}
where $\displaystyle {\rm ind}(L)=\frac{1}{3} \left(6 k_1 (1 + k_2^2) + k_2 (11 + k_2^2)\right)$. The different regions in $k$-space are shown in Fig.~\ref{fig:SecondMfd}.
\begin{figure}[h]
\begin{center}
{\hskip0pt
\begin{minipage}[t]{5.4in}
\raisebox{0in}{\includegraphics[width=7.2cm]{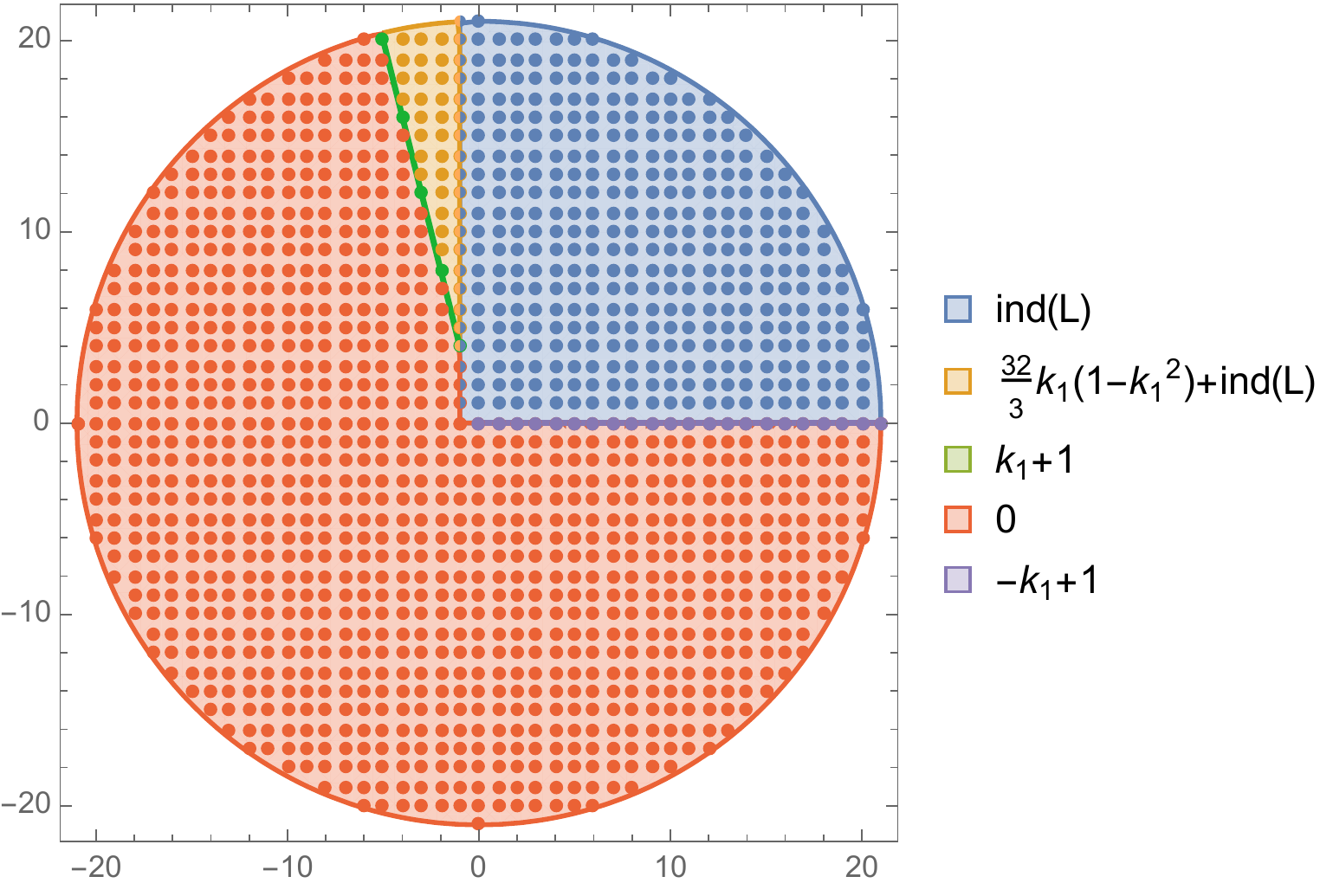}} 
\hfill \hspace*{14pt}
\raisebox{0in}{\includegraphics[width=7.2cm]{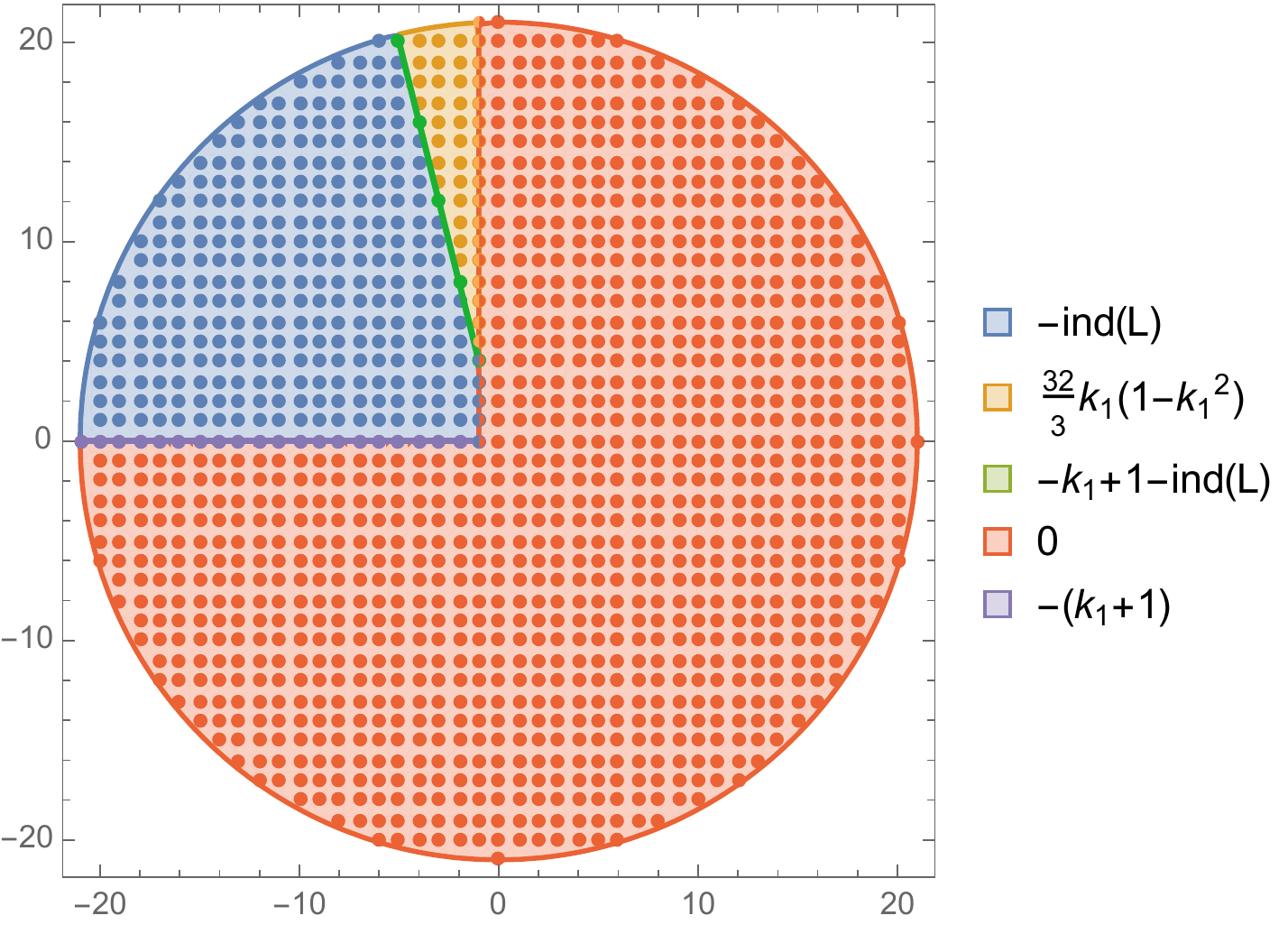}}
\hspace{10pt}
\vspace{10pt}
\end{minipage}}
\capt{6.4in}{fig:SecondMfd}{Regions in $k$-space where $h^0(X,L)$ (left) and $h^1(X,L)$ (right) have different expressions. In the blue regions $h^0(X,L) = {\rm ind}(L)$ and $h^1(X,L) = -{\rm ind}(L)$.}
\end{center}
\end{figure}

The above formulae can, in principle, be shown to hold in a way similar to the previous case of the bicubic manifold, that is, by starting with the sequence~\eqref{Koszul}. The novelty here (and also the feature which makes the proof more difficult) are the regions $k_1\leq -1$, $k_2=-4k_1$ and $k_1\leq -1$, $k_2>-4k_1$ which are  cones in the $k$-space whose tips are away from the origin. In these regions, the cohomology groups are given by 
\begin{equation*}
\begin{aligned}
H^0(X,L)&~\simeq~{\rm Ker}  \left( H^1(\cA,\cL \otimes \cN^*)  ~\longrightarrow~ H^1(\cA, \cL) \right)\\[4pt]
H^1(X,L)&~\simeq~{\rm Coker}  \left( H^1(\cA,\cL \otimes \cN^*)  ~\longrightarrow~ H^1(\cA, \cL) \right)\\[4pt]
H^2(X,L)&~\simeq ~H^3(X,L)~\simeq ~0
\end{aligned}
\end{equation*}
but the ranks of the maps involved in the expressions for $H^0(X,L)$ and $H^1(X,L)$ are non-maximal. 

\subsection{A co-dimension two manifold with Picard number two}
It is a reasonable question to ask whether the appearance of exact cohomology formulae is general, at least within the class of complete intersection Calabi-Yau manifolds, or merely an accidental phenomenon particular to certain manifolds. Without aiming at any kind of general proof, we can probe this question by studying a number of additional examples  with the aim of finding polynomial expressions of degree at most three for the ranks of all line bundle valued cohomology groups. 

The lesson learnt from the previous two examples is that relatively simple cohomology formulae appear irrespective of the details of the Koszul sequence. In particular, we have seen that details of how cohomologies on the ambient space relate to those on the Calabi-Yau sub-manifold or specific properties of the maps involved do not matter. The basic structure of the final result remains unchanged. 

The two previous examples had co-dimension one. Since the complexity of the calculation based on the sequence~\eqref{Koszul} increases significantly with increasing co-dimension it is reasonable to ask whether the same might hold for the final result. Perhaps surprisingly, the answer seems to be ``no''. Even at higher co-dimension, the final formula for the cohomology dimensions remains a cubic in the line bundle integers $k_r$, for each region in $k$-space. The purpose of this section is to illustrate this hypothesis with a co-dimension two complete intersection Calabi-Yau three-fold.  

Thus, let $X$ be a generic threefold of the family in the ambient space ${\cal A}=\mathbb{P}^2\times \mathbb{P}^3$ defined by the configuration matrix
\begin{equation*}
\cicy{\IP^2 \\ \IP^3}{\,2&\,1~ \\ \,2 &\,2~}^{2,62}~
\end{equation*}
and $L = \cO_X(k_1,k_2)$ a line bundle over $X$. It turns out that all explicit cohomology calculations for specific values of $k_1$, $k_2$ are consistent with the following formulae:
\begin{align}
h^0(X,L)  & = \begin{cases}\displaystyle \frac{1}{2} (1 + k_1) (2 + k_1)~, & k_1\geq0,~k_2=0 \\[8pt] 
{\rm ind}(L), & k_1\geq0,~k_2>0\\[8pt]
 8 k_1 (2 - 3 k_1^2) +{\rm ind}(L) ,~ & k_1<0~,k_2\geq-6k_1\\[8pt]
0&\text{otherwise}\end{cases}
\\[8pt]
h^1(X,L)  &= \begin{cases}\displaystyle\frac{1}{2}(1-k_1)(2-k_1)~, & k_1>0,~k_2=0\\[8pt]
{\rm Max}(-{\rm ind}(L),0)~, & k_1>0,~ -k_1+1< k_2<0\\[8pt]
{\rm Max}(-{\rm ind}(L),0)~ , & k_1 < 0,~ -k_1 - 1\leq k_2 < -6 k_1,~ k_2 > 0  \\[8pt] 
\displaystyle 8 k_1 (2 - 3 k_1^2) ~,& k_1 < 0,~ k_2 \geq -6 k_1\\[8pt] 
0 & \text{otherwise}\end{cases}
\end{align}
where ${\rm ind}(L)=\displaystyle\frac{1}{3} \left( 6 k_1^2 k_2 + 9 k_1 (1 + k_2^2) + k_2 (11 + k_2^2) \right)$. The regions associated to the case distinctions in the above formulae are shown in~\fref{fig:ThirdMfd}. Evidently, the structure of these results is quite similar to what we have seen for co-dimension one.
\begin{figure}[h]
\begin{center}
{\hskip0pt
\begin{minipage}[t]{5.4in}
\raisebox{0in}{\includegraphics[width=7.2cm]{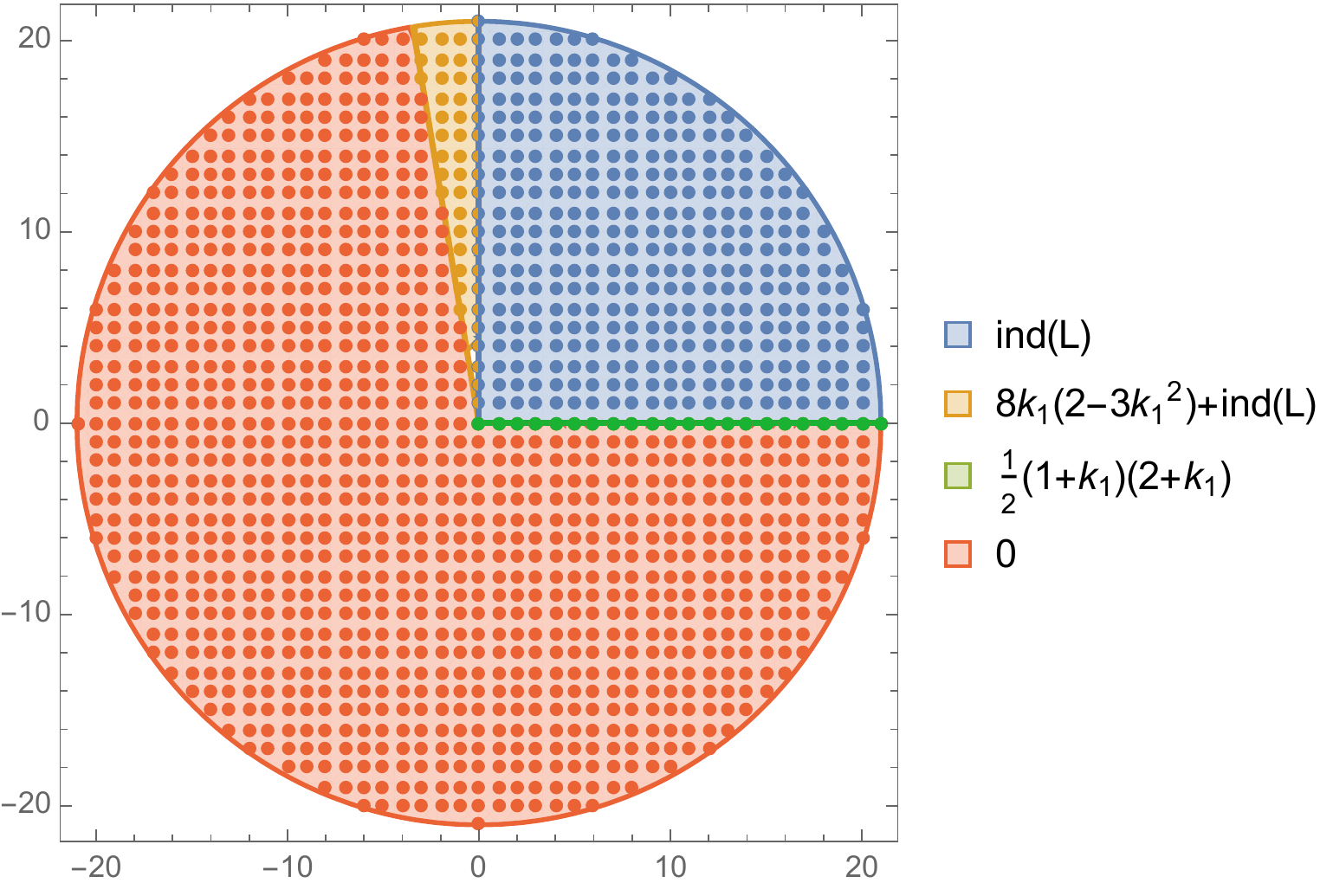}} 
\hfill \hspace*{14pt}
\raisebox{0in}{\includegraphics[width=7.2cm]{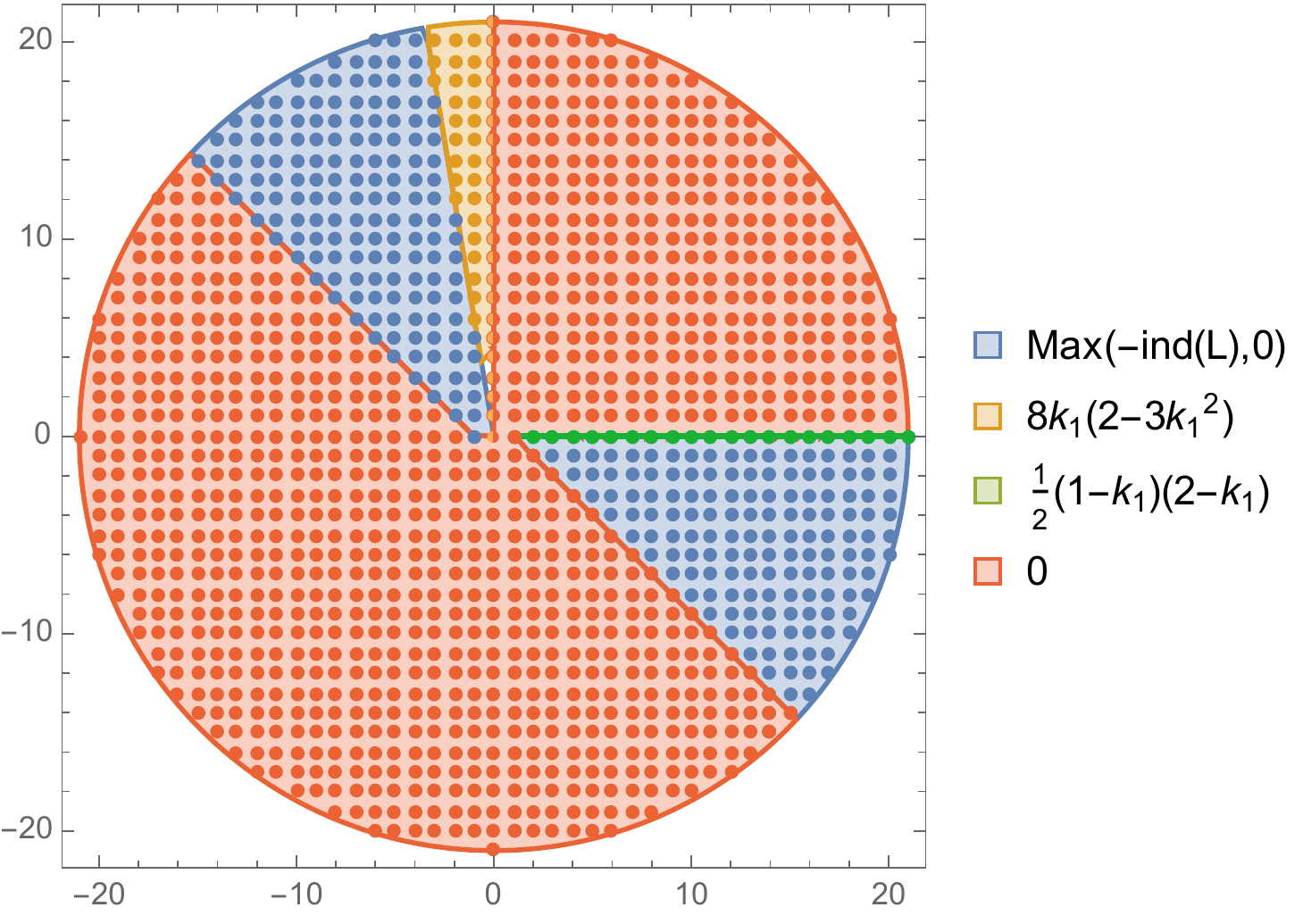}}
\hspace{10pt}
\end{minipage}}
\capt{5.8in}{fig:ThirdMfd}{Regions in $k$-space where $h^0(X,L)$ (left) and $h^1(X,L)$ (right) have different expressions. In the blue regions $h^0(X,L) = {\rm ind}(L)$ and $h^1(X,L) = -{\rm ind}(L)$. In the orange region the spectral sequence computation involves linear maps with non-maximal ranks.}
\vspace{-21pt}
\end{center}
\end{figure}

\subsection{A hypersurface with Picard number three}

The previous examples suggest that the general structure of the cohomology formulae is, as one would expect, insensitive to the realisation of $X$ as an embedding in a product of projective spaces and in particular to the co-dimension of~$X$. What is more important is the rank of ${\rm Pic}(X)$. As we will see, while the structure of the formulae remains unchanged for larger Picard numbers, the number of case distinctions increases. We illustrate this with two examples at Picard numbers three and four.

Let $X$ be a generic threefold in the family defined in the ambient space ${\cal A}=\mathbb{P}^1\times \mathbb{P}^1\times\mathbb{P}^2$ by the configuration matrix
\begin{equation*}
\cicy{\IP^1 \\ \IP^1 \\ \IP^2}{\,2~ \\ \,2~ \\ \,3~}^{3,75}~
\end{equation*}
and $L = \cO_X(k_1,k_2,k_3)$ a line bundle over $X$. Due to the symmetry between the two $\mathbb{P}^1$ factors, we can assume, without loss of generality, that $k_1\leq k_2$. 

The index of $L = \cO_X(k_1,k_2,k_3)$ is given by
\begin{equation*}
{\rm ind}(L) = (3k_1k_2 +k_1k_3+k_2k_3) k_3  +2(k_1+k_2)+3k_3~
\end{equation*}
and appears in various places below.
If $k_3=0$, the following cohomology formulae hold: 
\begin{align}
h^0(X,L)  & = \begin{cases}
(1+k_1)(1+k_2)~,& k_1\geq0,~k_2\geq0\\[8pt]
0&\text{otherwise}\end{cases}
\end{align}

\begin{align}
h^1(X,L)  &= \begin{cases}
(-1+k_1)(-1+k_2)~,&k_1>0,~k_2>0\\[12pt]
(-1- k1) (1+ k2)~,&k_1<0,~k_2\geq0\\[12pt]
0 & \text{otherwise}~.\end{cases}
\end{align}

If $k_3<0$, $h^0(L) = 0$ and
\begin{equation}
h^1(X,L) = \begin{cases} 
	{\rm Max}\left(-{\rm ind}(L),0\right)~,& k_1>0~,k_2>0\\[12pt]
	0~,& \text{otherwise}~.
	\end{cases}
\end{equation}

If $k_3>0$, and assuming again that $k_1\leq k_2$ we have: 
\begin{align}
h^0(X,L)  & = \begin{cases}
	\displaystyle\frac{1}{2}   (1 + k_3) (2 + k_3)~,& k_1=0,~k_2=0\\[12pt]
	{\rm ind}(L)~,& k_1\geq 0,~k_2>0\\[12pt]
	(1-k_1)(1+2k_1+k_2)~,& k_1 < 0,~ k_2 \geq -2 k_1-1,~ k_3 = -3 k_1 \\[12pt]
	9 k_1 (1 - k_1^2)+ {\rm ind}(L) ~,& k_1 < 0,~ k_2 \geq -2 k_1-1,~ k_3 > -3 k_1 \\[12pt]
	0&\text{otherwise}
	\end{cases}
\\[12pt]
h^1(X,L)  &= \begin{cases}
	\displaystyle\frac{1}{2}   (-1 + k_3) (-2 + k_3)~,& k_1=0,~k_2=0\\[12pt]
	{\rm Max}\left(-{\rm ind}(L),0\right)~,&k_1<0~,k_3<-3k_1\text{ or }k_1<0~,k_2<-2k_1-1\\[12pt]
\displaystyle (1 - k_1) (1 + 2 k_1 + k_2) - {\rm ind}(L) ~,&k_1 < -1~, k_2 \geq -2 k_1-1~, k_3 = -3 k_1\\[12pt] 
{\rm Max}\left(-{\rm ind}(L), 9 k_1 (1 - k_1^2)\right)~,&k_1 < -1~, k_2 \geq 0 ~, k_3 \neq -3 k_1\\[12pt]
0 & \text{otherwise}\end{cases}
\end{align}

While complicated, these formulae follow the same pattern as encountered in the earlier examples for manifolds with Picard numbers one and two: the $k$-space can be divided into several regions, and in each such region the ranks of bundle-valued cohomology groups can be expressed as polynomials in the $k$-integers of degree at most three. Note, however, that these regions are not cones in general. For instance, in the present example, when $k_1>0,k_2>0$ and $k_3<0$ we have $h^1(X,L)={\rm Max}\left(-{\rm ind}(L),0\right)$. This means that the region where $h^1(X,L) = -{\rm ind}(L)$ is bounded by the cubic surface ${\rm ind}(L)=0$.

\subsection{The tetraquadric manifold}

Let $X$ be the family of threefolds defined in the ambient space ${\cal A}=\mathbb{P}^1\times\mathbb{P}^1\times\mathbb{P}^1\times\mathbb{P}^1$ by the configuration matrix
\beq
\cicy{\IP^1 \\ \IP^1 \\ \IP^1\\ \IP^1}{\,2~ \\ \,2~ \\ \,2~\\ \,2~}^{4,68}~
\eeq
and $L = \cO_X(k_1,k_2,k_3,k_4)$ a line bundle over $X$, with index given by 
\begin{equation*}
{\rm ind}(L) = 2\left(k_1+k_2+k_3+k_4 \right) + 2\left( k_1 k_2 k_3 + k_1 k_2 k_4 + k_1 k_3 k_4 + k_2 k_3 k_4 \right)~.
\end{equation*}

For the tetraquadric manifold exact line bundle cohomology formulae have already appeared in Refs.~\cite{Constantin:2013, Buchbinder:2013dna}. Although equivalent, the formulae presented below are simpler and are expressed in terms of polynomials of degree at most $3$ in the line bundle integers. 

For the purpose of clarity, we assume, without loss of generality, that $k_1\leq k_2\leq k_3\leq k_4$ and, as before, we only present the formulae for $h^0(X,L)$ and $h^1(X,L)$, the other two cohomology groups being obtained by Serre duality. 

For $k_4<0$, Kodaira's vanishing theorem implies that $h^0(X,L)=h^1(X,L)=0$. Similarly, for $k_4=0$ we have 
\begin{align}
h^0(X,L)  &= \begin{cases}
1~,&k_1=k_2=k_3=0\\[12pt]
0 & \text{otherwise}\end{cases}
\\[12pt]
h^1(X,L)  &= \begin{cases}
-(1+k_1)~,&k_1<0,~k_2=k_3=0\\[12pt]
0 & \text{otherwise}~.\end{cases}
\end{align}

Hence we are left to discuss the case~$k_4>0$, which we do from now on.  

For $k_3<0$, $h^0(X,L)=h^1(X,L)=0$. For $k_3=0$, $h^0(X,L)=0$ and
\begin{align}
h^0(X,L)  &= \begin{cases}
(1+ k4)~,&k_1=k_2=0\\[8pt]
0 & \text{otherwise}\end{cases}
\\[8pt]
h^1(X,L)  &= \begin{cases}
-(1+k_1)(1+k_4)~,&k_1<0,~k_2=0\\[8pt]
0 & \text{otherwise}\end{cases}
\end{align}

From now on we assume that $k_3,k_4>0$. Then: 
\begin{align}
h^0(X,L)  & = \begin{cases}
	\displaystyle(1 + k_3) (1 + k_4)~,& k_1=0,~k_2=0\\[8pt]
	{\rm ind}(L)~,& k_1\geq 0,~k_2>0\\[8pt]
	{\rm ind}(L)-(1 + k_1) (-1 - 6 k_1 + 8 k_1^2 + k_4)~,& k_1 < 0,~ k_2 =k_3= -2 k_1,~ k_4 \geq -2 k_1 \\[8pt]
	{\rm ind}(L)-8 k_1 (-1 + k_1^2)~,& k_1 < 0,~ k_2 =-2k_1-1,~k_3,k_4> -2 k_1 ~\text{ or } \\[8pt]
		& k_1 < 0,~ k_2,k_3,k_4\geq -2 k_1 \\[8pt]
	{\rm Max}(0,{\rm ind}(L))&\text{otherwise}
	\end{cases}
\\[2pt]
h^1(X,L)  &= \begin{cases}
	\displaystyle(-1 + k_3) (-1 + k_4)~,& k_1=0,~k_2=0\\[8pt]
	0~,& k_1\geq 0,~k_2>0\\[8pt]
	{\rm Max}(0,-{\rm ind}(L))~,& k_1< 0,~k_2\leq 0\\[8pt]
	-(1 + k_1) (-1 - 6 k_1 + 8 k_1^2 + k_4)~,& k_1 < 0,~ k_2 =k_3= -2 k_1,~ k_4 \geq -2 k_1 \\[8pt]
	-8 k_1 (-1 + k_1^2)~,& k_1 < 0,~ k_2 =-2k_1-1,~k_3,k_4> -2 k_1 ~\text{ or } \\[8pt]
		& k_1 < 0,~ k_2,k_3,k_4\geq -2 k_1 \\[8pt]
	{\rm Max}(0,- {\rm ind}(L))& \text{otherwise}\end{cases}
\end{align}

These formulae have been checked to hold for all line bundles with integers in the range $-30\leq k_r\leq 30$.

\section{Non-simply connected Calabi-Yau threefolds}

We can enlarge the class of manifolds for which exact cohomology formulae can be studied by looking at non-simply connected Calabi-Yau threefolds realised as free quotients of complete intersections by discrete symmetries. 
More concretely, let $G$ be a finite group and $G\times X\rightarrow X$ a free holomorphic action of $G$ on the Calabi-Yau threefold $X$. Then the quotient $X/G$ is a smooth Calabi-Yau threefold with fundamental group isomorphic to $G$. Smooth quotients of complete intersection Calabi-Yau threefolds have been systematically studied in \cite{Candelas:2008wb, Braun:2010vc, Candelas:2010ve,Candelas:2015amz, Constantin:2016xlj} (see also the reviews \cite{Davies:2011fr, Candelas:2016fdy}). 

If $L\rightarrow X$ is a line bundle equivariant with respect to the action of $G$ on $X$, then $L$ is the pull-back of a line bundle on the quotient $X/G$. In the examples discussed below, all line bundles will be equivariant.

\subsection{A $\IZ_5$-quotient of the quintic threefold}

The quintic family $\IP^4[5]^{1,101}$ contains manifolds that admit a freely-acting $Z_5$-symmetry. Let $z_0$, $z_1$, $z_2$, $z_3$, $z_4$ be homogeneous coordinates on $\IP^4$ and consider the $\IZ_5$-action generated by 
\begin{equation*}
z_i \rightarrow \zeta^i z_i~,
\end{equation*}
where $\zeta$ is a non-trivial fifth root of unity. There are $26$ monomials invariant under this action. Let $X$ be a quintic manifold defined as the zero locus of a generic linear combination of these invariant monomials. Then $X$ admits a smooth quotient with Hodge numbers $(h^{1,1}(X),h^{2,1}(X))=(1,21)$. 

All line bundles $L=\cO_X(k)$ are equivariant with respect to the above action and we denote by $\tilde L$ the line bundle on $X/\IZ_5$ whose pullback is $L$. Then the following cohomology formulae hold:

\begin{equation}
h^0(X/\IZ_5,\tilde L) =~~ 
	\begin{cases} 
	{\rm ind}(\tilde L)~, &k>0\\[4pt] 
	1~, &k=0\\[4pt]
	0~, &k<0~.
	\end{cases}
\end{equation}	
Here, the index is given by ${\rm ind}(\tilde L) = \displaystyle \frac{1}{5}{\rm ind}(L) = \frac{1}{6} k^3 + \frac{10}{12} k$.

\subsection{A $\IZ_3$-quotient of the bicubic threefold}

The family of bicubic manifolds contains a sub-family which admits free quotients by a following $\IZ_3$-action. Introducing homogeneous coordinates $x_0,x_1,x_2$ and $y_0,y_1,y_2$ for the two ambient space $\mathbb{P}^2$ factors, the free $\IZ_3$-action is generated by 
\begin{equation*}
x_i\rightarrow \omega^i x_i~,~~~ y_i\rightarrow \omega^i y_i~,
\end{equation*}
with $\omega$ a non-trivial cube root of unity. There are $34$ monomials invariant under the above action and we consider a cubic manifold $X$ defined by a generic linear combination of the invariant monomials. By quotienting, a Calabi-Yau threefold with Hodge numbers $(2,29)$ and fundamental group $\IZ_3$ is obtained.

As in the previous example, all line bundles $L=\cO_X(k_1,k_2)$ are equivariant with respect to the above $\IZ_3$-action. We denote by $\tilde L$ the bundle on $X/\IZ_3$ whose pullback is $L$. Then the following cohomology formulae hold:

\begin{align}
h^0(X/\IZ_3,L)  &  = 
	\begin{cases}
	\displaystyle \frac{1}{6}(1+k_2)(2+k_2)~, & k_1=0,~k_2\geq 0,~ k_2 = 1~{\rm mod }~3~{\rm or }~k_2 = 2~{\rm mod }~3\\[12pt]
	\displaystyle \frac{k_2^2}{6}+\frac{k_2}{2}+1~, & k_1=0,~k_2\geq 0,~ k_2 = 0~{\rm mod }~3\\[12pt] 
	\displaystyle\frac{1}{2}(k_1+k_2)(2+k_1k_2)~, & k_1,k_2>0\\[12pt]
	0 & \text{otherwise}
	\end{cases}
\end{align}
\begin{align}
h^1(X/\IZ_3,L)  & = 
	\begin{cases}
	\displaystyle \frac{1}{6} (-1+ k_2) (-2 + k_2) ~, & k_1=0,~k_2>0~, k_2 = 1~{\rm mod }~3~{\rm or }~k_2 = 2~{\rm mod }~3\\[12pt] 
	\displaystyle \frac{k_2^2}{6}-\frac{k_2}{2}+1~, & k_1=0,~k_2> 0,~ k_2 = 0~{\rm mod }~3\\[12pt] 
	\displaystyle-\frac{1}{2} (k_1 + k_2) (2 + k_1 k_2)~, & k_1<0,~k_2>-k_1\\[12pt]
	0 & \text{otherwise}~.
	\end{cases}
\end{align}

\section{Conclusions}
The evidence gathered from the examples presented in this note suggests that the existence of relatively simple formulae for line bundle cohomology on Calabi-Yau three-folds is a generic phenomenon. We have studied several complete intersection Calabi-Yau threefolds and some of their quotients  by freely acting discrete symmetries, with Picard numbers ranging from one to four and we have found a common pattern. The space of line bundle integers can be divided into several different regions, and in each such region the ranks of all cohomology groups can be expressed as a polynomial in the line bundle integers of degree at most three. 

The formulae presented here were found by computing cohomology dimensions for a large number of line bundles and by looking for patterns in this data. The computations were carried out using a Mathematica implementation of a computational algorithm that relies on the Bott-Borel-Weil theorem and spectral sequences techniques applied to the Koszul sequence~\eqref{Koszul}. Although conceptually straightforward, these calculations are often laborious and computationally intensive. Despite the intricacy of the intermediate computations and the presence of Leray maps with non-maximal ranks, the final cohomology results fall into the simple pattern described above. It is relatively straightforward to fix the cubic polynomials which describe the cohomology dimensions by matching to sufficiently many line bundle cohomologies. The more difficult part of extracting the correct formulae from the data is to establishing the regions of validity for these polynomials.

For each Calabi-Yau manifold that we have analysed, the cohomology dimensions have been computed for line bundles with integers in the range $10\leq k_r\leq 10$. For each manifold, their number is significantly larger than what is required in order to fix all the coefficients in the general Ansatz for the cohomology formula, yet all cohomology results are correctly described by this formula. While this is of course not a proof it provides a non-trivial check of our results.

For relatively simple cases the cohomology formulae can be proved by chasing through the long exact cohomology sequences associated to the Koszul sequence~\eqref{Koszul} and, where required, computing ranks of maps using methods of commutative algebra. We have carried this out explicitly for the bi-cubic in $\mathbb{P}^2\times\mathbb{P}^2$. For more complicated examples this approach, while possible in principle, becomes extremely cumbersome and it would not be practical to carry this out even for a modest number of manifolds. 

There are two other potential ways of deriving or extracting cohomology formulae in a systematic way. For our examples, the structure of the formulae turns out to be independent of the ambient space. This suggest that there may be an alternative, more intrinsic method to compute these cohomologies which bypasses the embedding into the ambient space and works on the Calabi-Yau manifold only. We do not currently know how such a method would work - or if it even exists - but it would certainly be interesting to pursue this further.

From a practical point of view, our results suggest a very concrete problem in machine learning. Such techniques have recently been applied to problems in geometry and string theory and for the pioneering papers see Refs.~\cite{He:2017aed,Ruehle:2017mzq, Krefl:2017yox, Carifio:2017bov, Bull:2018uow, Demirtas:2018akl}. We can use the line bundle cohomology data on a given manifold, as computed by the methods described in Ref.~\cite{cicypackage}, and train a neural network. However, unlike for most applications of machine learning, the goal would not merely be to have the neural network predict further cohomology results for individual line bundles, but rather to extract concrete formulae from the trained neural network. Such an approach would facilitate extracting cohomology formulae in a systematic way and for a large number of manifolds. Work in this direction is currently underway. 

After this paper appeared, Ref.~\cite{Klaewer:2018sfl} was submitted to the arXiv. This work discusses related problems in the context of Calabi-Yau hypersurfaces in toric four-folds, extracting information about line bundle cohomology using machine learning techniques. The basic structure of their results is similar to the one presented here. 

\section*{Acknowledgements}
We are grateful to James Gray, Damian R\"ossler and Shing-Tung Yau  for insightful discussions. A.~C.~would like to thank the University of Oxford for hospitality during part of the completion of this project. A.~L.~is partially supported by the EPSRC network grant EP/N007158/1.
%


\begin{thebibliography}{10}

\bibitem{Blumenhagen:2010pv}
R.~Blumenhagen, B.~Jurke, T.~Rahn, and H.~Roschy, ``{Cohomology of Line
  Bundles: A Computational Algorithm},'' {\em J. Math. Phys.} {\bf 51} (2010)
  103525,
\href{http://arXiv.org/abs/1003.5217}{{\tt 1003.5217}}.

\bibitem{Rahn:2010fm}
T.~Rahn and H.~Roschy, ``{Cohomology of Line Bundles: Proof of the
  Algorithm},'' {\em J. Math. Phys.} {\bf 51} (2010) 103520,
\href{http://arXiv.org/abs/1006.2392}{{\tt 1006.2392}}.

\bibitem{Jow:2010}
S.-Y. Jow, ``{Cohomology of Toric Line Bundles via Simplicial Alexander
  Duality},'' \href{http://arXiv.org/abs/1006.0780}{{\tt 1006.0780}}.

\bibitem{Candelas:1987kf}
P.~Candelas, A.~Dale, C.~Lutken, and R.~Schimmrigk, ``{Complete Intersection
  Calabi-Yau Manifolds},'' {\em Nucl.Phys.} {\bf B298} (1988)
493.

\bibitem{Candelas:1987du}
P.~Candelas, C.~A. Lutken, and R.~Schimmrigk, ``{Complete Intersection
  Calabi-Yau Manifolds 2. Three Generation Manifolds},'' {\em Nucl. Phys.} {\bf
  B306} (1988)
113.

\bibitem{bestiary}
T.~H\"ubsch, {\em Calabi-Yau Manifolds: A Bestiary for Physicists}.
\newblock World Scientific, 1991.

\bibitem{Anderson:2008ex}
L.~B. Anderson, ``{Heterotic and M-theory Compactifications for String
  Phenomenology},''
\href{http://arXiv.org/abs/0808.3621}{{\tt 0808.3621}}.

\bibitem{cicypackage}
L.~B. Anderson, J.~Gray, Y.-H. He, S.-J. Lee, and A.~Lukas, ``{CICY package",
  based on methods described in arXiv:0911.1569, arXiv:0911.0865,
  arXiv:0805.2875, hep-th/0703249, hep-th/0702210},''.

\bibitem{Anderson:2007nc}
L.~B. Anderson, Y.-H. He, and A.~Lukas, ``{Heterotic Compactification, An
  Algorithmic Approach},'' {\em JHEP} {\bf 0707} (2007) 049,
\href{http://arXiv.org/abs/hep-th/0702210}{{\tt hep-th/0702210}}.

\bibitem{Anderson:2008uw}
L.~B. Anderson, Y.-H. He, and A.~Lukas, ``{Monad Bundles in Heterotic String
  Compactifications},'' {\em JHEP} {\bf 0807} (2008) 104,
\href{http://arXiv.org/abs/0805.2875}{{\tt 0805.2875}}.

\bibitem{Anderson:2009mh}
L.~B. Anderson, J.~Gray, Y.-H. He, and A.~Lukas, ``{Exploring Positive Monad
  Bundles And a New Heterotic Standard Model},'' {\em JHEP} {\bf 1002} (2010)
  054,
\href{http://arXiv.org/abs/0911.1569}{{\tt 0911.1569}}.

\bibitem{Anderson:2011ns}
L.~B. Anderson, J.~Gray, A.~Lukas, and E.~Palti, ``{Two Hundred Heterotic
  Standard Models on Smooth Calabi-Yau Threefolds},'' {\em Phys.Rev.} {\bf D84}
  (2011) 106005,
\href{http://arXiv.org/abs/1106.4804}{{\tt 1106.4804}}.

\bibitem{Anderson:2012yf}
L.~B. Anderson, J.~Gray, A.~Lukas, and E.~Palti, ``{Heterotic Line Bundle
  Standard Models},'' {\em JHEP} {\bf 1206} (2012) 113,
\href{http://arXiv.org/abs/1202.1757}{{\tt 1202.1757}}.

\bibitem{Constantin:2013}
A.~Constantin, ``{Heterotic String Models on Smooth Calabi-Yau Threefolds
  (DPhil Thesis)},'' {\em ${\rm Oxford~U.}$} (2013), \href{http://arXiv.org/abs/1808.09993}{{\tt 1808.09993}}.

\bibitem{Buchbinder:2013dna}
E.~I. Buchbinder, A.~Constantin, and A.~Lukas, ``{The Moduli Space of Heterotic
  Line Bundle Models: a Case Study for the Tetra-Quadric},'' {\em JHEP} {\bf
  1403} (2014) 025,
\href{http://arXiv.org/abs/1311.1941}{{\tt 1311.1941}}.

\bibitem{Anderson:2013xka}
L.~B. Anderson, A.~Constantin, J.~Gray, A.~Lukas, and E.~Palti, ``{A
  Comprehensive Scan for Heterotic SU(5) GUT models},'' {\em JHEP} {\bf 1401}
  (2014) 047,
\href{http://arXiv.org/abs/1307.4787}{{\tt 1307.4787}}.

\bibitem{He:2013ofa}
Y.-H. He, S.-J. Lee, A.~Lukas, and C.~Sun, ``{Heterotic Model Building: 16
  Special Manifolds},''
\href{http://arXiv.org/abs/1309.0223}{{\tt 1309.0223}}.

\bibitem{Buchbinder:2014qda}
E.~I. Buchbinder, A.~Constantin, and A.~Lukas, ``{A heterotic standard model
  with $B - L$ symmetry and a stable proton},'' {\em JHEP} {\bf 1406} (2014)
  100,
\href{http://arXiv.org/abs/1404.2767}{{\tt 1404.2767}}.

\bibitem{Buchbinder:2014sya}
E.~I. Buchbinder, A.~Constantin, and A.~Lukas, ``{Non-generic Couplings in
  Supersymmetric Standard Models},'' {\em Phys. Lett.} {\bf B748} (2015)
  251--254,
\href{http://arXiv.org/abs/1409.2412}{{\tt 1409.2412}}.

\bibitem{Buchbinder:2014qca}
E.~I. Buchbinder, A.~Constantin, and A.~Lukas, ``{Heterotic QCD axion},'' {\em
  Phys. Rev.} {\bf D91} (2015), no.~4, 046010,
\href{http://arXiv.org/abs/1412.8696}{{\tt 1412.8696}}.

\bibitem{Anderson:2014hia}
L.~B. Anderson, A.~Constantin, S.-J. Lee, and A.~Lukas, ``{Hypercharge Flux in
  Heterotic Compactifications},'' {\em Phys. Rev.} {\bf D91} (2015), no.~4,
  046008,
\href{http://arXiv.org/abs/1411.0034}{{\tt 1411.0034}}.

\bibitem{Constantin:2015bea}
A.~Constantin, A.~Lukas, and C.~Mishra, ``{The Family Problem: Hints from
  Heterotic Line Bundle Models},'' {\em JHEP} {\bf 03} (2016) 173,
\href{http://arXiv.org/abs/1509.02729}{{\tt 1509.02729}}.

\bibitem{Buchbinder:2016jqr}
E.~I. Buchbinder, A.~Constantin, J.~Gray, and A.~Lukas, ``{Yukawa Unification
  in Heterotic String Theory},'' {\em Phys. Rev.} {\bf D94} (2016), no.~4,
  046005,
\href{http://arXiv.org/abs/1606.04032}{{\tt 1606.04032}}.

\bibitem{Braun:2017feb}
A.~P. Braun, C.~R. Brodie, and A.~Lukas, ``{Heterotic Line Bundle Models on
  Elliptically Fibered Calabi-Yau Three-folds},''
\href{http://arXiv.org/abs/1706.07688}{{\tt 1706.07688}}.

\bibitem{Blesneag:2018ygh}
S.~Blesneag, E.~I. Buchbinder, A.~Constantin, A.~Lukas, and E.~Palti, ``{Matter
  field Kähler metric in heterotic string theory from localisation},'' {\em
  JHEP} {\bf 04} (2018) 139,
\href{http://arXiv.org/abs/1801.09645}{{\tt 1801.09645}}.

\bibitem{Klaus:2018}
K.~Altmann, J.~Buczynski, L.~Kastner, and A.-L. Winz, ``{Immaculate line
  bundles on toric varieties},''
\href{http://arXiv.org/abs/1808.09312}{{\tt 1808.09312}}.

\bibitem{Candelas:2008wb}
P.~Candelas and R.~Davies, ``{New Calabi-Yau Manifolds with Small Hodge
  Numbers},'' {\em Fortsch.Phys.} {\bf 58} (2010) 383--466,
\href{http://arXiv.org/abs/0809.4681}{{\tt 0809.4681}}.

\bibitem{Braun:2010vc}
V.~Braun, ``{On Free Quotients of Complete Intersection Calabi-Yau
  Manifolds},'' {\em JHEP} {\bf 1104} (2011) 005,
\href{http://arXiv.org/abs/1003.3235}{{\tt 1003.3235}}.

\bibitem{Candelas:2010ve}
P.~Candelas and A.~Constantin, ``{Completing the Web of $Z_3$ - Quotients of
  Complete Intersection Calabi-Yau Manifolds},'' {\em Fortsch.Phys.} {\bf 60}
  (2012) 345--369,
\href{http://arXiv.org/abs/1010.1878}{{\tt 1010.1878}}.

\bibitem{Candelas:2015amz}
P.~Candelas, A.~Constantin, and C.~Mishra, ``{Hodge Numbers for CICYs with
  Symmetries of Order Divisible by 4},'' {\em Fortsch. Phys.} {\bf 64} (2016),
  no.~6-7, 463--509,
\href{http://arXiv.org/abs/1511.01103}{{\tt 1511.01103}}.

\bibitem{Constantin:2016xlj}
A.~Constantin, J.~Gray, and A.~Lukas, ``{Hodge Numbers for All CICY
  Quotients},'' {\em JHEP} {\bf 01} (2017) 001,
\href{http://arXiv.org/abs/1607.01830}{{\tt 1607.01830}}.

\bibitem{Davies:2011fr}
R.~Davies, ``{The Expanding Zoo of Calabi-Yau Threefolds},'' {\em Adv. High
  Energy Phys.} {\bf 2011} (2011) 901898,
\href{http://arXiv.org/abs/1103.3156}{{\tt 1103.3156}}.

\bibitem{Candelas:2016fdy}
P.~Candelas, A.~Constantin, and C.~Mishra, ``{Calabi-Yau Threefolds with Small
  Hodge Numbers},'' {\em Fortsch. Phys.} {\bf 66} (2018), no.~6, 1800029,
\href{http://arXiv.org/abs/1602.06303}{{\tt 1602.06303}}.

\bibitem{He:2017aed}
Y.-H. He, ``{Deep-Learning the Landscape},''
\href{http://arXiv.org/abs/1706.02714}{{\tt 1706.02714}}.

\bibitem{Ruehle:2017mzq}
F.~Ruehle, ``{Evolving neural networks with genetic algorithms to study the
  String Landscape},'' {\em JHEP} {\bf 08} (2017) 038,
\href{http://arXiv.org/abs/1706.07024}{{\tt 1706.07024}}.

\bibitem{Krefl:2017yox}
D.~Krefl and R.-K. Seong, ``{Machine Learning of Calabi-Yau Volumes},'' {\em
  Phys. Rev.} {\bf D96} (2017), no.~6, 066014,
\href{http://arXiv.org/abs/1706.03346}{{\tt 1706.03346}}.

\bibitem{Carifio:2017bov}
J.~Carifio, J.~Halverson, D.~Krioukov, and B.~D. Nelson, ``{Machine Learning in
  the String Landscape},'' {\em JHEP} {\bf 09} (2017) 157,
\href{http://arXiv.org/abs/1707.00655}{{\tt 1707.00655}}.

\bibitem{Bull:2018uow}
K.~Bull, Y.-H. He, V.~Jejjala, and C.~Mishra, ``{Machine Learning CICY
  Threefolds},''
\href{http://arXiv.org/abs/1806.03121}{{\tt 1806.03121}}.

\bibitem{Demirtas:2018akl}
M.~Demirtas, C.~Long, L.~McAllister, and M.~Stillman, ``{The Kreuzer-Skarke
  Axiverse},''
\href{http://arXiv.org/abs/1808.01282}{{\tt 1808.01282}}.

\bibitem{Klaewer:2018sfl}
D.~Klaewer, and L.~Schlechter, ``{Machine Learning Line Bundle Cohomologies of Hypersurfaces in Toric Varieties},''
\href{http://arXiv.org/abs/1809.02547}{{\tt 1809.02547}}.


\end{thebibliography}

\bibliographystyle{utcaps}
\providecommand{\href}[2]{#2}\begingroup\raggedright\endgroup

\end{document}